\documentclass[preprint,prc,superscriptaddress,unsortedaddress,showpacs,preprintnumbers,amsmath,amssymb]{revtex4}
\usepackage{tipa}

\usepackage{graphicx}
\usepackage{dcolumn}
\usepackage{bm}
\usepackage{color}

\def\beq{\begin{equation}}
\def\eeq{\end{equation}}
\def\bea{\begin{eqnarray}}
\def\eea{\end{eqnarray}}

\def\fun#1#2{\lower3.6pt\vbox{\baselineskip0pt\lineskip.9pt
  \ialign{$\mathsurround=0pt#1\hfil##\hfil$\crcr#2\crcr\sim\crcr}}}

\preprint{}

\begin{document}

\title{Nucleon momentum distributions in asymmetric nuclear matter}

\author{Z. X. Yang}
 \affiliation{Institute
of Modern Physics, Chinese Academy of Sciences, Lanzhou 730000,
China}\affiliation{School of Nuclear Science and Technology,
University of Chinese Academy of Sciences, Beijing 100049, China}

\author{X. L. Shang}\email[ ]{shangxinle@impcas.ac.cn}
 \affiliation{Institute
of Modern Physics, Chinese Academy of Sciences, Lanzhou 730000,
China}\affiliation{School of Nuclear Science and Technology,
University of Chinese Academy of Sciences, Beijing 100049, China}
\author{G. C. Yong}
 \affiliation{Institute
of Modern Physics, Chinese Academy of Sciences, Lanzhou 730000,
China}\affiliation{School of Nuclear Science and Technology,
University of Chinese Academy of Sciences, Beijing 100049, China}

\author{W. Zuo}
\affiliation{Institute of Modern Physics, Chinese Academy of
Sciences, Lanzhou 730000, China}\affiliation{School of Nuclear
Science and Technology, University of Chinese Academy of Sciences,
Beijing 100049, China}

\author{Y. Gao}
 \affiliation{School of Information Engineering,
 Hangzhou Dianzi University, Hangzhou 310018, China}

\begin{abstract}
Nucleon momentum distributions at various densities and
isospin-asymmetries for nuclear matter are investigated
systematically within the extended Bruecker-Hartree-Fock approach.
The shapes of the normalized momentum distributions varying with
$k/k_{F}$ are practically identical, while the density and isospin
dependent magnitude of the distribution is directly related to the
depletion of the Fermi sea. Based on these properties, a
parameterized formula is proposed with the parameters calibrated to
the calculated result.

\end{abstract}


\maketitle

\section{INTRODUCTION}
To determine reliably the structure and properties of nuclear matter
is one of the central issues in nuclear physics and nuclear
astrophysics \cite{sp1,sp2,sp3,sp4}. One of the most important
properties of nuclear matter is the neutron and proton momentum
distributions which can shed light on the correlations between
nucleons \cite{md1,md2,md3,md4}. In an ideal infinite noninteracting
Fermi systems at zero temperature, the momentum distribution is the
step function, i.e., $n(k)=\theta(k_{F}-k)$, and the Fermi sea is
fully occupied. Once the interactions are turned on, the
correlations induced by the interactions among fermions lead to the
occupation of states with momenta $k>k_{F}$ (the high-momentum
distribution) and the depletion of the Fermi sea \cite{lm1,lm2,lm3}.
In addition, the depletion can be straightly obtained from the
momentum distribution. As for the nuclear matter, due to the hard
core and the tensor component of the $NN$ interaction, the depletion
of the Fermi sea is quite significant \cite{nlm1,nlm2}. It measures
the dynamical $NN$ correlation strength induced by the $NN$
interaction \cite{md3}, and is believed to be an indicator for
testing the validity of physical picture of independent particle
motion in the mean field approach or standard shell model
\cite{ipm1,ipm2} in a nuclear many-body system. The knowledge of the
momentum distribution in nuclear matter may provide useful
information to acquaint the depletion of the deeply bound state
inside finite nuclei and then to understand the structure beyond
mean field theory of finite nuclei. It may as well help ones to
study the effects of short-range correlations (SRCs) on the
observables in the heavy-ion reactions \cite{yong1,yong2}.

Experimentally, the high-momentum distribution and the $NN$
correlations were unambiguously identified in series of experiments,
such as $(e,e^{'}p)$ \cite{eht1}, $(e,e^{'}NN)$ \cite{nn1,nn2} and
so on. Especially, the two-nucleon knockout experiment shows that
nucleons can form short-range correlated pairs with large relative
momenta and small center-of-mass momenta \cite{srcp2}. The number of
neutron-proton (np) correlated pairs was found to be about 18 times
that of proton-proton (pp) correlated pairs \cite{srcp1,nn3,nn4}
which suggests that the tensor correlations due to the strong tensor
components of the $NN$ interaction, in addition to SRCs, play also
an important role in the high-momentum distributions \cite{np1}. In
Ref. \cite{md4}, the authors attempted to distinguish the dominant
regions of tensor correlations and SRCs via comparing the momentum
distributions of nuclear matter with the deuteron. They found that
SRCs tend to dominate the high-momentum distributions above $3 \ \
\texttt{fm}^{-1}$, while the tensor correlations is of interest in
the region of $k \sim 2-2.8 \ \ \texttt{fm}^{-1}$. However, both in
the theoretical calculations and experiments, the effects of these
two correlations on the momentum distribution are hard to
distinguished strictly.

In theoretical calculations, the $NN$ correlations in nuclear matter
have often been studied in combination with the nucleon momentum
distribution. Various theoretical methods have been employed to
study these distributions, such as the correlated basis functions
\cite{cbf1,cbf2}, quantum Monte Carlo method \cite{qmc}, the
self-consistent Green's function (SCGF)
\cite{md4,scg1,scg2,scg3,scg4}, the in-medium T -matrix method
\cite{tm1,tm2} and the extended Brueckner-Hartree-Fock (EBHF) method
\cite{md1,bhf1,bhf2,bhf3,bhf4}. In Ref. \cite{scg4}, the
temperature, density, and isospin dependence of the depletion of the
Fermi sea is clarified in the framework of SCGF. The momentum
distribution at large momentum has been discussed as well which
shows an exponential damping tendency. In Ref. \cite{bhf4}, the
authors have calculated the nucleon momentum distribution and
quasiparticle strength in symmetric nuclear matter within EBHF
approach. Parameterized three-section expression of the momentum
distribution fit to the microscopic calculation has also been
provided. Unfortunately, the parametrization is density independent
and merely valid for symmetric case. In the present paper, we shall
extend the parameterized expression of the momentum distribution to
asymmetric nuclear matter and simplify the form of the expression of
the momentum distribution. Moreover, the density and isospin
dependence of the depletion of the Fermi sea is discussed as well
within the EBHF approach. In order to obtain a more realistic
momentum distribution expression, the calculated momentum
distribution within the EBHF approach includes the three-body force
(TBF) effects.

This paper is organized as follows. In the next section, we give a
brief review of the adopted theoretical approaches including the
EBHF theory and spectral function. The formula of the momentum
distribution is derived in Sec. III. In Sec. IV, we employ the
obtained formula to study the SRC effects on the heavy-ion
reactions. And finally, a summary is given in Sec. V.


\section{THEORETICAL APPROACHES}
The present calculations for asymmetric nuclear matter are based on
the EBHF approach, for which one can refer to Ref. \cite{EBHF} for
details. The extension of the BHF scheme to include microscopic TBF
can be found in Refs. \cite{tbf1,tbf2}. After several
self-consistent iteration, the effective interaction matrix G in the
Brueckner-Bethe-Goldstone (BBG) theory can be obtained. This
G-matrix which include all the ladder diagrams of the $NN$
interaction embodies the tensor correlations and the SRCs. Using the
G-matrix, the mass operator $M(k,\omega)$ can be calculated.

\subsection{The mass operator within the extended Brueckner-Hatree-Fock approach }
Generally, the nucleon momentum distribution needs the exact
knowledge of the mass operator. In practice, it is impossible to
calculate the mass operator exactly. In an actual calculation, one
can only evaluate some approximations to the mass operator. Within
the framework of the BBG theory, the mass operator can be expanded
in a perturbation series according to the number of hole lines. To
the lowest-order approximation, i.e., the BHF approximation, the
mass operator is written as
\begin{eqnarray}
M_{1}(k,\omega)=\sum_{k'}\theta(k_{F}-k')\langle
kk'|G[\omega+\epsilon(k')]|kk'\rangle_{A},
\end{eqnarray}
where $\omega$ is the starting energy and $\epsilon(k)$ represents
the s.p. spectrum in the BHF approximation. The step function
$\theta(k_{F}-k)$ is the Fermi distribution at zero temperature. The
subscript $A$ denotes antisymmetrization of the matrix elements.

The quantity $M_{1}(k,\omega)$ only has a right-hand cut which is
mainly responsible for the depletion under Fermi surface
\cite{lm3,nlm1}. Therefore, the calculation of the momentum
distribution requires at least the first two order approximation of
the mass operator. The second order in the hole-line expansion of
the mass operator, which might be answerable for the high-momentum
distributions above Fermi surface \cite{lm3}, is given by
\cite{EBHF}
\begin{eqnarray}
M_{2}(k,\omega)&=&\frac{1}{2}\sum_{k'k_{1}k_{2}}\theta(k'-k_{F})\theta(k_{F}-k_{1})\theta(k_{F}-k_{2})\nonumber\\&\times&\frac{|\langle
kk'|G[\epsilon(k_{1})+\epsilon(k_{2})]|k_{1}k_{2}\rangle_{A}|^{2}}{\omega+\epsilon(k')-\epsilon(k_{1})-\epsilon(k_{2})-i0},
\end{eqnarray}
where the step function $\theta(k'-k_{F})$ guarantees the integral
over $k'$ above the Fermi surface. In the present paper, we
calculate the mass operator to the second order approximation, i.e.,
$M(k,\omega)\cong M_{1}(k,\omega)+M_{2}(k,\omega)$.

\subsection{The spectral function and the momentum distribution}
The knowledge of $M(k,\omega)$ allows us to write down the Green's
function in the energy-momentum representation,
\begin{eqnarray}
\mathcal
{G}(k,\omega)=\frac{1}{\omega-\frac{k^{2}}{2m}-M(k,\omega)}.
\end{eqnarray}
Except at the Fermi energy $\epsilon_{F}$, the mass operator
$M(k,\omega)$ is complex and can be written as
\begin{eqnarray}
M(k,\omega+i\eta)=V(k,\omega)+i W(k,\omega)
\end{eqnarray}
with the property $[M(k,\omega+i\eta)]^{*}=M(k,\omega-i\eta)$, where
$\eta=+0$ to ensure the integral-path. The spectral function
$S(k,\omega)$, which describes the probability density of removing a
particle with momentum k from a target nuclear system and leaving
the final system with the excitation energy $\omega$, is thus given
by
\begin{eqnarray}
S(k,\omega)&=&\frac{i}{2\pi}[\mathcal {G}(k,\omega)-\mathcal
{G}(k,\omega)^{*}]\nonumber\\
&=&-\frac{1}{\pi}\frac{W(k,\omega)}{[\omega-k^2/2m-V(k,\omega)]^{2}+W(k,\omega)^{2}}.\nonumber\\
\end{eqnarray}
And it should fulfill the sum rule
\begin{eqnarray}
\int_{-\infty}^{\infty}S(k,\omega)d\omega=1.
\end{eqnarray}
In Ref. \cite{md2}, the authors show that an elaborately dealing
with the integral over the energy can satisfy the sum rule quite
well by adopting the mass operator up to second order in the
framework of EBHF approach.

Finally, the momentum distribution $n(k)$ is related to the
spectral function by
\begin{eqnarray}
n(k)=\int_{-\infty}^{\epsilon_{F}}S(k,\omega)d\omega
\end{eqnarray}
or equivalently,
\begin{eqnarray}
n(k)=1-\int_{\epsilon_{F}}^{\infty}S(k,\omega)d\omega.
\end{eqnarray}
The Fermi energy $\epsilon_{F}$ follows the on shell condition
$\epsilon_{F}=k_{F}^2/2m+V(k,\epsilon_{F})$. Using the momentum
distribution $n(k)$, one can then define the depletion parameter
\begin{eqnarray}
\chi=[\sum_{\textbf{k}}n(k>k_{F})]/\rho
=[\frac{1}{\pi^{2}}\int_{k_{F}}^{\infty}n(k)k^{2}dk]/\rho,
\end{eqnarray}
i.e., the proportion of the particle number above the Fermi
momentum. Which is related to several physical quantities such as
the correlation strength or the defect function, and is believed to
be an indicator for the convergence of the so-called BBG hole-line
expansion.


\section{THE FORMULA OF THE MOMENTUM DISTRIBUTION}
In this section, we first exhibit the numerical calculation of
momentum distributions within the EBHF approach, then roughly
analyze the behavior of these distributions, and finally provide a
formula of calculating the distribution. The realistic Argonne $V18$
two-body interaction supplemented with a microscopic 3BF
\cite{tbf1,tbf2} is taken as the $NN$ interaction. In the present
paper, the calculation is under zero temperature.

\begin{figure}
\includegraphics[scale=0.30]{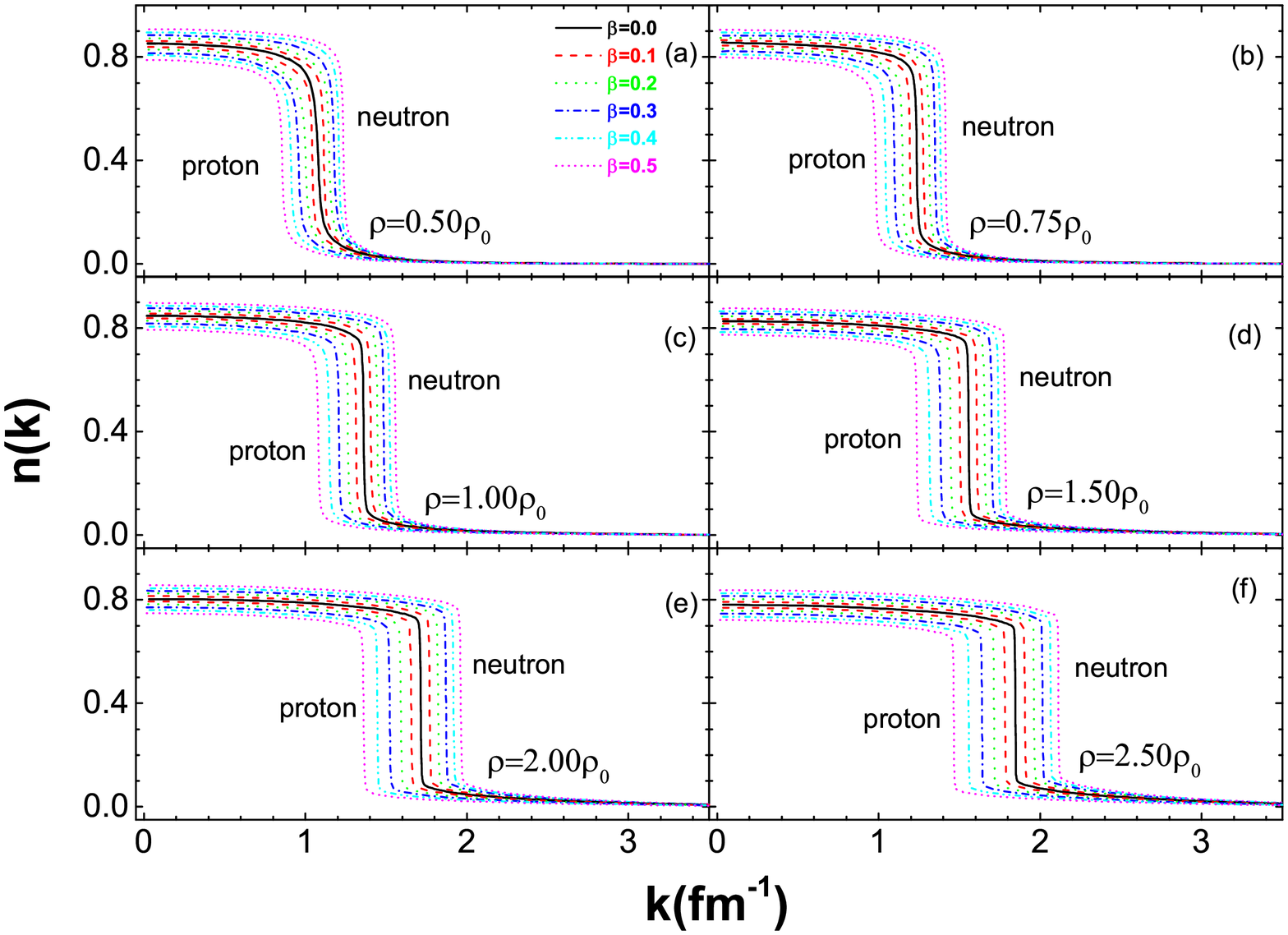} \caption{(Color online).
Neutron and proton momentum distributions in asymmetric nuclear
matter at various isospin-asymmetries calculated within the EBHF
approach. Fig. (a)$\sim$(f) represent the different densities
0.50$\rho_{0}$, 0.75$\rho_{0}$, 1.0$\rho_{0}$, 1.50$\rho_{0}$,
2.0$\rho_{0}$, 2.5$\rho_{0}$} \label{bhfdata}
\end{figure}
We systematically report the calculated neutron and proton momentum
distributions at various isospin-asymmetries $\beta=$0.0, 0.1, 0.2,
0.3, 0.4, 0.5 with different total densities 0.50$\rho_{0}$,
0.75$\rho_{0}$, 1.0$\rho_{0}$, 1.50$\rho_{0}$, 2.0$\rho_{0}$,
2.5$\rho_{0}$ in Fig. 1. Hereafter, the isospin-asymmetry $\beta$ is
defined as $\beta=(\rho_ {n}-\rho_{p})/(\rho_ {n}-\rho_{p})$ and
$\rho_{0}=$ 0.17 $\texttt{fm}^{-3}$ is the empirical saturation
density of nuclear matter. The distributions present a discontinuity
at their respective Fermi momenta $k_{F}^{\tau}$ (hereafter
$\tau=n,p$). For positive asymmetries, the neutron Fermi momentum
$k_{F}^{n}$ is larger than the proton Fermi momentum $k_{F}^{p}$,
therefore the proton and neutron momentum distributions are located
at the left and right sides of the symmetric case, respectively. One
should note that the neutron momentum distribution differs only
slightly from proton momentum distribution in symmetric case due to
the charge-dependent interaction Argonne $V18$. The discrepancy is
too tiny to be recognized in the Figs.

\begin{figure}
\includegraphics[scale=0.30]{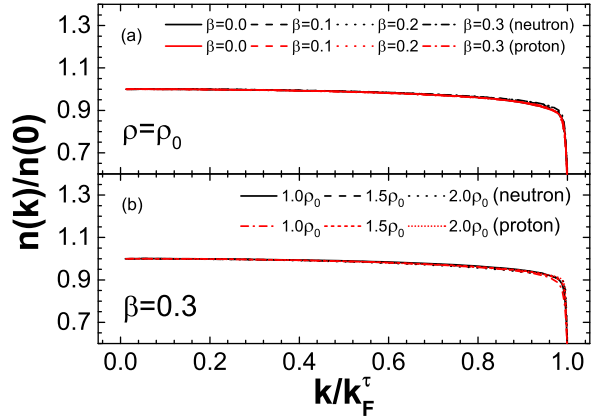} \caption{(Color online).
The normalized momentum distribution $n(k)/n(0)$ vs $k/k_{F}^{\tau}$
below the Fermi momentum at various isospin-asymmetries (upper
panel) and at various densities (lower pannel).} \label{lowd}
\end{figure}

\begin{figure}
\includegraphics[scale=0.30]{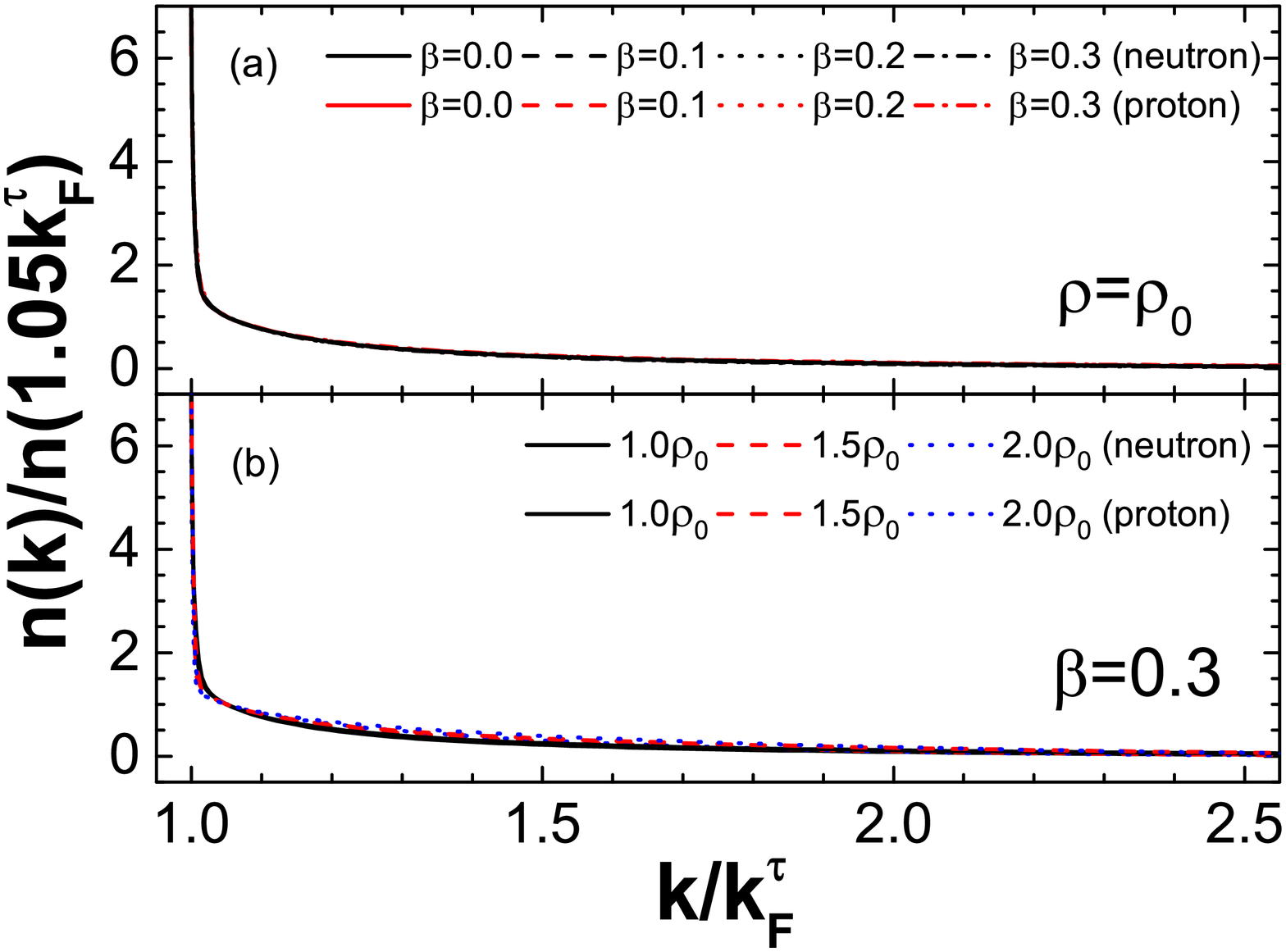} \caption{(Color online).
The normalized momentum distribution $n(k)/n(1.05k_{F})$ vs
$k/k_{F}$ above the Fermi momentum at various isospin-asymmetries
(upper panel) and various densities (lower pannel).} \label{upd}
\end{figure}
Interestingly, if focusing on the shapes of the momentum
distributions in Fig.1, one would notice that these distributions
are quite similar except the magnitudes. Inspired by this quality,
we show the normalized momentum distributions $n(k)/n(0)$ as a
function of the ratio $k/k_{F}^{\tau}$ below the Fermi momentum at
various isospin asymmetries and densities in Fig. 2. Where
$k_{F}^{\tau}$ is the respective Fermi momentum corresponding to the
different isospin-asymmetry $\beta$ and density $\rho$. The shapes
of the normalized distributions are practically identity except
slightly small discrepancies near the Fermi momentum. In Fig. 3, the
same normalized momentum distributions as Fig. 2 but above the Fermi
moment exhibit the coincidence of the shapes as well. In other
words, the normalized momentum distributions as a function of the
ratio $k/k_{F}^{\tau}$ below (above) the Fermi momentum at various
densities and isospin-asymmetries can be described by the same
expression with tolerable errors.

\begin{figure}
\includegraphics[scale=0.30]{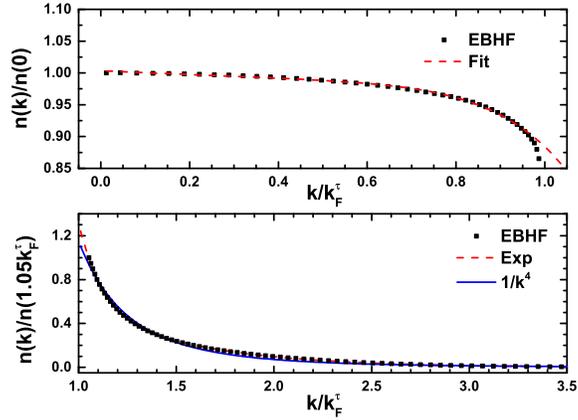} \caption{(Color online).
The normalized distributions as a function of $k/k_{F}$ and the
fittings. The upper and the lower panels correspond to the
momentum below and above the momentum, respectively.}
\label{taild}
\end{figure}
In the domain $0<k/k_{F}^{\tau}<1$, the normalized momentum
distribution varying with $k/k_{F}^{\tau}$ below the Fermi momentum
can be described by the following parametrization:
\begin{eqnarray}
\frac{n_{<}^{\tau}(k)}{n(0)}=1.00329-0.02876x-0.09053x^{7},
\end{eqnarray}
where $n_{<}^{\tau}(k)$ corresponds to $n(k<k_{F}^{\tau})$ and $x$
represents the ratio $k/k_{F}^{\tau}$. A comparison between the
calculated $n_{<}^{\tau}(k)/n(0)$ and the parametrization is shown
in the upper panel of Fig. 4. The polynomial fit is in good
agreement with the calculation within the EBHF approach.

For the high-momentum distributions, i.e., $k>k_{F}^{\tau}$, Ref.
\cite{k4} reports that the momentum distributions appear to decrease
as $k^{-4}$, following the Tan's relation \cite{tan1,tan2}. However,
the Tan's relation is simply valid for dilute system with contact
interaction whereas the $NN$ interaction is much more complicated.
And the microscopic calculations including the EBHF approach
\cite{bhf4} and the SCGF method \cite{md4,scg4} indicate a nearly
linear relation between $\ln n(k)$ and $k$ at large momentum, i.e.,
$n(k)\propto\exp(-ck)$ ($c$ is a positive constant). In addition, if
one adopts the form of $k^{-4}$ to describe the high-momentum
distributions, a cutoff $k_{\Lambda}$ is always supplemented owing
to the slow convergence of the number density. When $k_{\Lambda}$ is
employed, the neglect number density is
\begin{eqnarray}
\int_{k_{\Lambda}}^{\infty}n(k)k^{2}dk\propto\frac{1}{k_{\Lambda}} \
\ .
\end{eqnarray}
In calculations, the maximum value of $k_{\Lambda}$ is usual about
$5\ \ \texttt{fm}^{-1}$, which implies three to five percent missing
of the number density. Most importantly, our calculations within the
EBHF approach reveal the same behavior of the high-momentum
distributions as Ref. \cite{bhf4}. On account of the above reasons,
we employ the exponential form replenished by a Gauss function to
describe the high-momentum behavior. The normalized momentum
distribution above the Fermi momentum can be expressed as
\begin{eqnarray}
\frac{n_{>}^{\tau}(k)}{n(1.05k_{F}^{\tau})}=3.548e^{-1.799x}+52.2e^{-4.2766x^{2}},
\end{eqnarray}
with $n_{>}^{\tau}(k)\equiv n(k>k_{F}^{\tau})$ and $x\equiv
k/k_{F}^{\tau}$. We display the expression of $k^{-4}$ ($1/k^{4}$),
the parametrization (12) (Exp) and the calculation within EBHF
approach in the lower panel of Fig. 4. Obviously, the exponential
fit is more approaching to the calculation than the $k^{-4}$ fit.
But we should stress that owing to the approximations adopted in the
calculations and fittings, the possibility of Tan's relation in
nucleon momentum distribution could not be ruled out.

\begin{figure}
\includegraphics[scale=0.30]{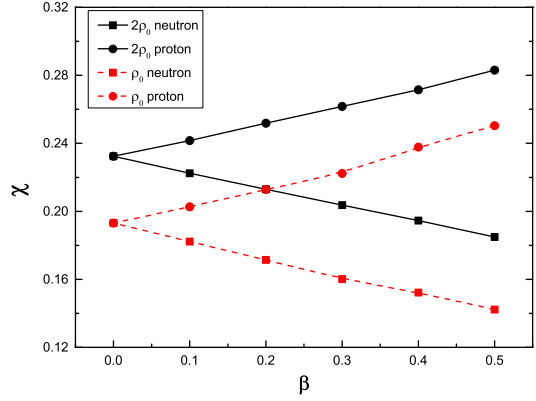} \caption{(Color online).
The depletion parameter $\chi$ calculated within EBHF approach
varying with isospin-asymmetry $\beta$ for two densities $\rho_{0}$
and 2.0$\rho_{0}$.} \label{asy}
\end{figure}

\begin{figure}
\includegraphics[scale=0.30]{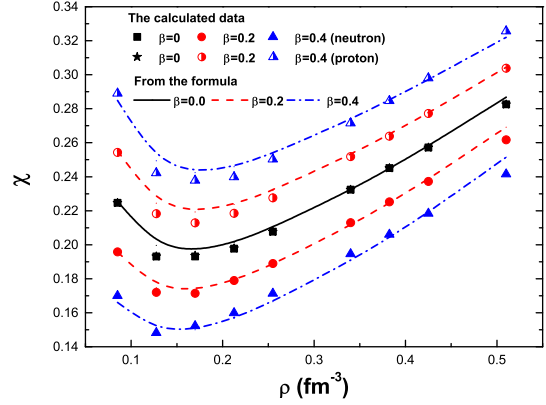} \caption{(Color online).
The depletion parameter $\chi$ vs densities. The symbols
correspond to the calculations within the EBHF approach. The lines
are obtained form the Eq. (16). The upper and lower lines are
related to proton and neutron, respectively.} \label{den}
\end{figure}
To obtain the momentum distributions, the magnitudes of $n(k)$ below
and above the Fermi momentum remain to be identified once the shapes
are provided. Actually, the magnitudes connect with the depletion
parameter $\chi$ via the relations
\begin{eqnarray}
1-\chi&=&\frac{\frac{1}{\pi^{2}}\int_{0}^{k_{F}^{\tau}}n_{<}^{\tau}(k)k^{2}dk}{\rho_{\tau}}\nonumber\\
&=&3\int_{0}^{1}n_{<}^{\tau}(k)x^{2}dx=0.9546n(0),\\
\chi&=&\frac{\frac{1}{\pi^{2}}\int_{k_{F}^{\tau}}^{\infty}n_{>}^{\tau}(k)k^{2}dk}{\rho_{\tau}}\nonumber\\
&=&3\int_{1}^{\infty}n_{>}^{\tau}(k)x^{2}dx=2.9537n(1.05k_{F}^{\tau}).
\end{eqnarray}
Consequently, the magnitudes of the $n(k)$ below and above the Fermi
momentum directly related to the depletion parameter, i.e.,
\begin{eqnarray}
n(0)=\frac{1-\chi}{0.9546},\  \ n(1.05k_{F})=\frac{\chi}{2.9537}\ \
.
\end{eqnarray}

The depletion parameter $\chi$ calculated within the EBHF approach
with various isospin-asymmetries $\beta=$ 0.0, 0.1, 0.2, 0.3, 0.4
and 0.5 at two typical densities $\rho_{0}$ and 2.0$\rho_{0}$ are
exhibited in Fig. 5. Obviously, the proton (neutron) depletion of
the Fermi sea increases (decreases) almost linearly with varying
isospin-asymmetry. Due to Eq. (15), $n(0)$ emerges the analogous
behavior which have been reported in Ref. \cite{md1,lin}. The
experiments show that the np correlation is much stronger than nn or
pp correlation \cite{nn3,nn4}. One should notice that the
probability of a proton (neutron) encounters a neutron (proton)
increases (decreases) linearly as a function of isospin-asymmetry.
If supposing equal correlation in each correlated np pair and
neglecting the nn/pp correlation, the linear isospin dependence of
$\chi$ comes very naturally. In Fig. 6 we illustrate the density
dependence of the depletion parameter. The different types of dots
correspond to the calculated $\chi$ within the EBHF approach.
According to the shapes of $\chi$ varying with $\rho$ , we propose
an expression with the parameters calibrated to the calculated
$\chi$. The expression reads
\begin{eqnarray}
\chi(\rho,\beta)&=&0.1669[1+\lambda(0.1407\frac{\rho}{\rho_{0}}-0.7296)\beta]\nonumber\\
&\times&[1+2.448e^{-4.1854\frac{\rho}{\rho_{0}}}+0.1382(\frac{\rho}{\rho_{0}})^{1.5}] \ \ .\nonumber\\
\end{eqnarray}
Where $\lambda=$ 1/-1 corresponds to neutron/proton. The isospin and
density dependence of $\chi$ are mainly included in the first and
second square brackets on the right-hand side of
 Eq. (16), respectively. One would find that there is a slight discrepancy
between the slopes of curves in Fig. 5 indicating a weak density
dependence of $\partial\chi/\partial\beta$. We actually account this
dependence in the first square bracket of Eq. (16). In fact, a
simple analysis of the calculated data on the density dependence of
the slope reveals a roughly linear dependence. In Ref. \cite{md4},
the authors have also mentioned a similar behavior of momentum
distribution in asymmetric nuclear matter at finite temperature. The
expression (16) for various densities and isospin-asymmetries are
shown by lines in Fig .6. Below the saturation density, the
depletion of the Fermi sea becomes stronger with decreasing density
which might mainly result from the increasing effect of the tensor
correlation. While above the saturation density, the hard-core
effect and the depletion get larger and large with increasing
density \cite{md3}.

\begin{figure}
\includegraphics[scale=0.30]{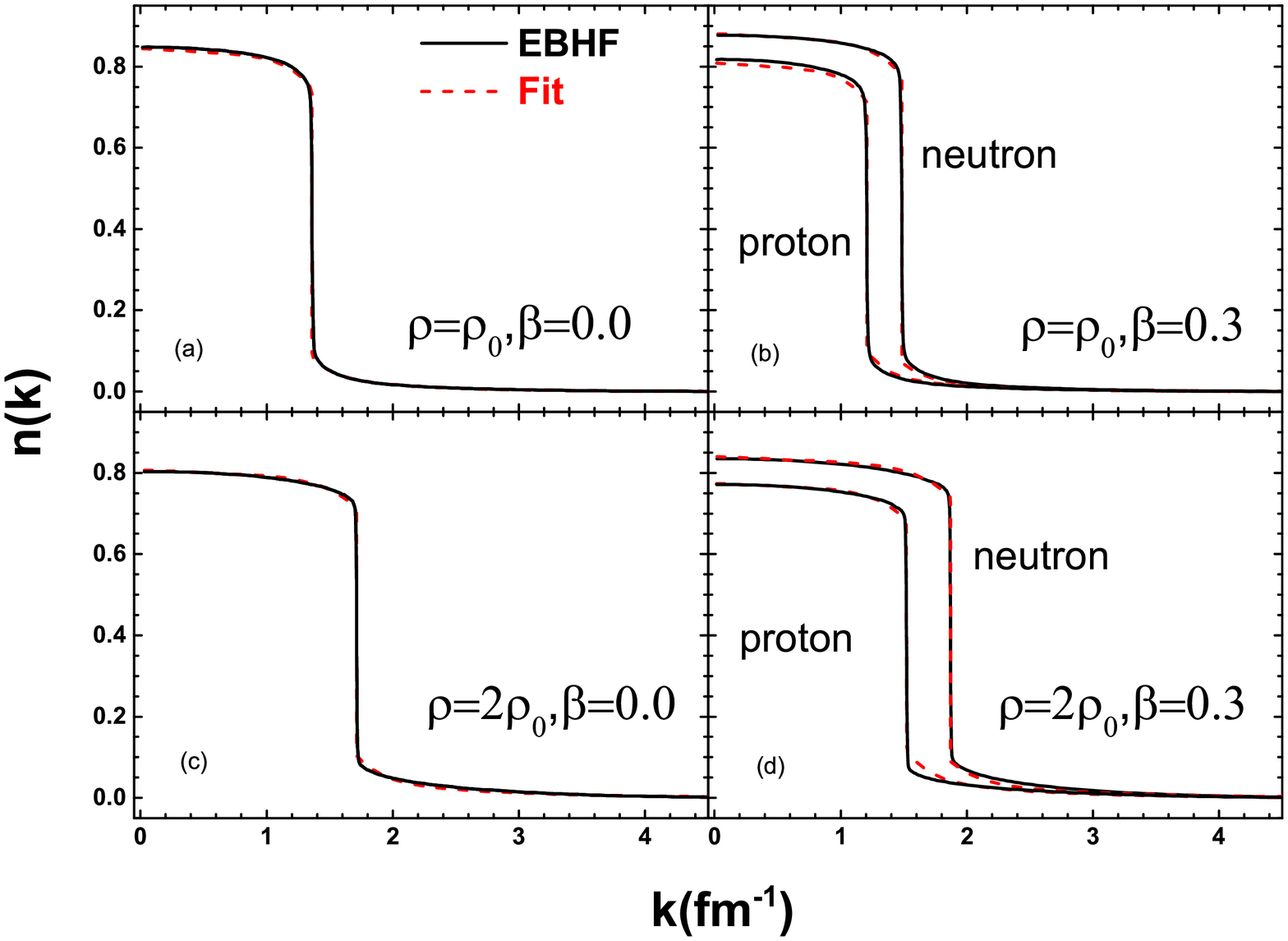} \caption{(Color online).
Neutron and proton momentum distributions form the formula (17) and
the calculation within the EBHF approach at two isospin-asymmetries
and densities.} \label{comp}
\end{figure}
Finally, the formula of the momentum distribution can be summarized
as
\begin{eqnarray}
n(k)=\begin{cases} \frac{1-\chi}{0.9546}[1.00329-0.02876\frac{k}{k_{F}^{\tau}}-0.09053(\frac{k}{k_{F}^{\tau}})^{7}]\\ \text{if} \ \ k\leq k_{F}^{\tau} \ \ ;\\
\frac{\chi}{2.9537}[3.548e^{-1.799\frac{k}{k_{F}^{\tau}}}+52.2e^{-4.2766(\frac{k}{k_{F}^{\tau}})^{2}}]\\
\text{if} \ \ k\geq k_{F}^{\tau} \ \ ,
\end{cases}
\end{eqnarray}
with the expression of the depletion parameter Eq. (16). A
comparison between the momentum distributions from formula (17) and
from the EBHF approach is given in Fig. 7. It can be clearly seen
that the formula is quite accurate except a slight difference near
the Fermi momentum. This formula can be applied to calculate the
momentum distribution in finite nuclei assisted by the local density
approximation. As is well known, at low densities the nuclear matter
system can minimize its energy by forming light cluster such as
deuterons, or particularly strongly bound $\alpha$ particle
\cite{clus}. In theoretical calculations such as EBHF approach, the
in-medium T-matrix method and SCGF method, the effective interaction
including all the ladder-diagram contribution always encounters a
singularity leading to unstable results at low densities
\cite{ust1,ust2}. Therefore, we emphasize that the achieved formula
(17) of the momentum distribution might be solely reliable for the
density of $0.1\rho_{0}<\rho <3.0\rho_{0}$ and the isospin-asymmetry
of $\beta\in(-0.5,0.5)$ for uniform nuclear matter. Otherwise, one
should be careful of the depletion parameter.


\section{Application to the transport model}

\begin{figure}[th]
\centering
\includegraphics[height=15cm,width=0.97\textwidth]{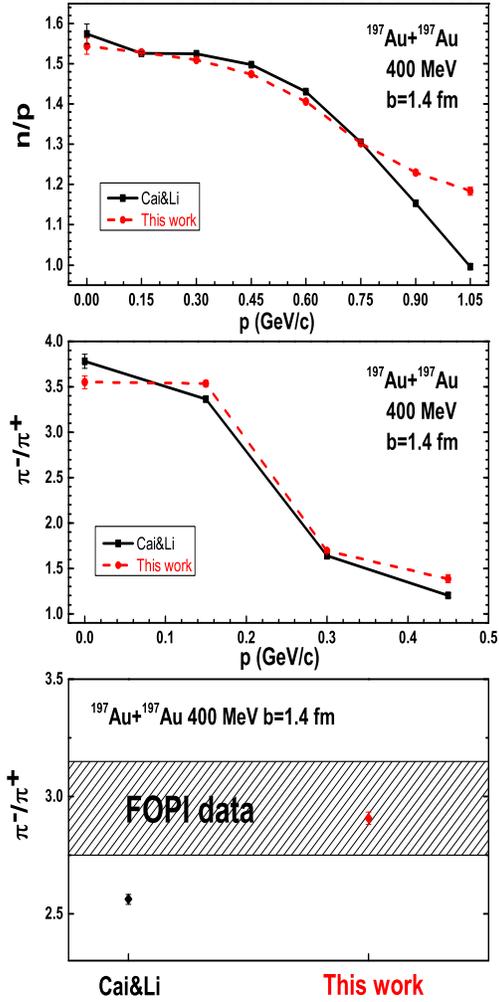}
\caption{(Color online) Upper panel: Free neutron to proton ratio as
a function of momentum in the central Au+Au reaction at 400
MeV/nucleon. Middle panel: Same as upper panel, but for
$\pi^{-}/\pi^{+}$ ratio. Lower panel: Comparison of calculated total
$\pi^{-}/\pi^{+}$ yields ratios and FOPI data \cite{fopi}.}
\label{rnpion}
\end{figure}
As an example of application, the obtained density and
asymmetry-dependent nucleon momentum distribution Eq.~(17) and the
fraction of high-momentum nucleons Eq.~(16) were both involved in
the isospin-dependent Boltzmann-Uehling-Uhlenbeck (IBUU) transport
model \cite{yonggc17}. The free neutron to proton ratio and the
$\pi^{-}/\pi^{+}$ ratio as a function of momentum in the central
Au+Au reaction at 400 MeV/nucleon are demonstrated in the upper and
middle windows of Fig.~\ref{rnpion}. As comparisons, nucleon
momentum distribution from Cai \& Li is also used
\cite{caibj,yong2}. The lower window shows the integrations of the
$\pi^{-}/\pi^{+}$ ratio and comparison with the FOPI data
\cite{fopi}. From Fig.~\ref{rnpion}, with the formula (17) and the
work of Cai \& Li, it is seen that both the momentum distribution of
the free n/p ratio and the $\pi^{-}/\pi^{+}$ ratio are quite
similar, except for energetic n/p ratio. While the difference of the
total $\pi^{-}/\pi^{+}$ yields ratios is evidently shown in the
lower window of Fig.~\ref{rnpion}. The value of $\pi^{-}/\pi^{+}$
yields ratio with the formula (17) is higher than that with the form
of Cai \& Li. The reason is that the fraction of high-momentum
nucleons with the formula (17) is smaller than that with the form of
Cai \& Li \cite{caibj} and larger number of neutron-proton
correlation causes a lower value of the $\pi^{-}/\pi^{+}$ ratio in
heavy-ion collisions.

\section{SUMMARY AND OUTLOOK}
Summarily, we have systematically calculated the nucleon momentum
distributions and the depletions of the Fermi sea at various
densities and isospin-asymmetries for nuclear matter within the EBHF
approach. The identity of these shapes of the normalized momentum
distributions below (above) the Fermi momentum varying with
$k/k_{F}$ is detected, indicating an uniform expression of the of
the momentum distribution for different densities and
isospin-asymmetries. Whereas the magnitude of the momentum
distribution is directly related to the depletion of the Fermi sea,
which first decreases and then increases with densities resulting
from the tensor and hard core effects of the $NN$ interaction
\cite{md3}. Using these properties, the parameterized formula of
momentum distribution is proposed with the expression of the
depletion.  Moreover, a heavy-ion reaction example adopting the
obtained formula is given to test its reliability.

In the present paper, the mass operator is just calculated up to the
second order, the missing higher order perhaps enhances the
depletion of the Fermi sea and eventually influences the momentum
distribution. Especially, the missing higher order might as well
reduce the particle strength around Fermi surface \cite{sh1}. Thus,
the parameterized formula cannot be considered as definite. In
addition, the calculation is based on realistic Argonne $V18$ only.
With different interactions, the depletions of the Fermi sea and the
momentum distributions would differ from each other
\cite{md3,md4,scg4}. Furthermore, the normal state of symmetric
nuclear matter becomes unstable owing to pairing tendency of np
\cite{ust1,ust2,sh1,sh2} and one should account the effect of the
pairing on the momentum distribution. An improvement of the
calculations including these effects is under way .


\section*{Acknowledgments}
{The work is supported by National Natural Science Foundation of
China (No. 11435014, 11505241, 11875013, 11775276, 11775275,
11705240), the Youth Innovation Promotion Association of Chinese
Academy of Sciences.}
\appendix


\begin{thebibliography}{90}

\vspace{3mm}

\bibitem{sp1}
B. A. Li, L.W. Chen, and C. M. Ko, Phys. Rep. {\bf 464}, 113 (2008).
\bibitem{sp2}
I. Bombaci and U. Lombardo, Phys. Rev. C {\bf 44}, 1892 (1991).
\bibitem{sp3}
A. Ramos, A. Polls and W. H. Dickhoff, Nucl. Phys. A {\bf 503}, 1
(1989).
\bibitem{sp4}
M. Baldo and H. R. Moshfegh, Phys. Rev. C {\bf 86}, 024306 (2012).
\bibitem{md1}
P. Yin, J. Y. Li, P. Wang, and W. Zuo, Phys. Rev. C {\bf 87}, 014314
(2013).
\bibitem{md2}
P. Wang, and W. Zuo, Phys. Rev. C {\bf 89}, 054319 (2014).
\bibitem{md3}
Z. H. Li and H.-J. Schulze, Phys. Rev. C {\bf 94}, 024322 (2016).
\bibitem{md4}
A. Rios, A. Polls and W. H. Dickhoff, Phys. Rev. C {\bf 89}, 044303
(2014).
\bibitem{lm1}
A. B. Migdal, Sov. Phys. -JETP {\bf 5}, 333 (1957) [Zh. Eksp. Teor.
Fiz. (USSR) {\bf 32} 339 (1957)].
\bibitem{lm2}
J. M. Luttinger, Phys. Rev. {\bf 119}, 1153 (1960).
\bibitem{lm3}
C. Mahaux and R. Sartor, Phys. Rep. {\bf 211}, 53 (1992).
\bibitem{nlm1}
J. P. Jeukenne, A. Lejeune, and C. Mahaux, Phys. Rep. {\bf 25}, 83
(1976).
\bibitem{nlm2}
B. E. Vonderfecht, W H. Dickhoff, A. Polls, and A. Ramos, Phys. Rev.
C {\bf 44}, R1265 (1991); Nucl. Phys. A {\bf 555}, 1 (1993).
\bibitem{ipm1}
V. R. Pandharipande, I. Sick, and P. K. A. deWitt Huberts, Rev. Mod.
Phys. {\bf 69}, 981 (1997).
\bibitem{ipm2}
J. M. Cavedon, B. Frois, D. Goutte et al., Phys. Rev. Lett. {\bf
49}, 978 (1982).
\bibitem{yong1}
G. C. Yong and B. A. Li, Phys. Rev. C {\bf 96}, 064614 (2017).
\bibitem{yong2}
Z. X. Yang, X. H. Fan, G. C. Yong and W. Zuo, Phys. Rev. C {\bf 98},
014623 (2018).
\bibitem{eht1}
D. Rohe et al., Phys. Rev. Lett. {\bf 93}, 182501 (2004).
\bibitem{nn1}
C. J. G. Onderwater et al., Phys. Rev. Lett. {\bf 81}, 2213 (1998).
\bibitem{nn2}
R. Starink et al., Phys. Lett. B {\bf 474}, 33 (2000).
\bibitem{srcp2}
R. Shneor et al., Phys. Rev. Lett. {\bf 99}, 072501 (2007).
\bibitem{srcp1}
E. Piasetzky, M. Sargsian, L. Frankfurt, M. Strikman, and J. W.
Watson, Phys. Rev. Lett. {\bf 97}, 162504 (2006).
\bibitem{nn3}
R. Subedi et al., Science {\bf 320}, 1476 (2008).
\bibitem{nn4}
O. Hen et al. (The CLAS Collaboration), Science {\bf 346}, 614
(2014).
\bibitem{np1}
R. Schiavilla, R. B. Wiringa, S. C. Pieper, and J. Carlson, Phys.
Rev. Lett. {\bf 98}, 132501 (2007).
\bibitem{cbf1}
S. Fantoni and V. Pandharipande, Nucl. Phys. A {\bf 427}, 473
(1984).
\bibitem{cbf2}
O. Benhar, A. Fabrocini, and S. Fantoni, Nucl. Phys. A {\bf 505},
267 (1989); Phys. Rev. C {\bf 41}, R24 (1990).
\bibitem{qmc}
A. Gezerlis and J. Carlson, Phys. Rev. C {\bf 81}, 025803 (2010).
\bibitem{scg1}
Y. Dewulf, D. Van Neck, and M. Waroquier, Phys. Rev. C {\bf 65},
054316 (2002); Y. Dewulf, W. H. Dickhoff, D. Van Neck, E. R.
Stoddard, and M. Waroquier, Phys. Rev. Lett. {\bf 90}, 152501
(2003).
\bibitem{scg2}
T. Frick, H.M¡§uther, A. Rios, A. Polls, and A. Ramos, Phys. Rev. C
{\bf 71}, 014313 (2005).
\bibitem{scg3}
A. Rios, A. Polls, and I. Vidana, Phys. Rev. C {\bf 79}, 025802
(2009)
\bibitem{scg4}
A. Rios, A. Polls, and W. H. Dickhoff, Phys. Rev. C {\bf 79}, 064308
(2009).
\bibitem{tm1}
P. Bozek, Phys. Rev. C {\bf 59}, 2619 (1999); {\bf 65}, 054306
(2002).
\bibitem{tm2}
V. Soma and P. Bozek, Phys. Rev. C {\bf 78}, 054003 (2008).
\bibitem{bhf1}
R. Sartor and C. Mahaux, Phys. Rev. C {\bf 21}, 1546 (1980); P.
Grang\'{e}, J. Cugnon, and A. Lejeune, Nucl. Phys. A {\bf 473}, 365
(1987); M. Jaminon and C. Mahaux, Phys. Rev. C {\bf 41}, 697 (1990);
M. Baldo, I. Bombaci, G. Giansiracusa, and U. Lombardo, Nucl. Phys.
A {\bf 530}, 135 (1991); C. Mahaux and R. Sartor, Nucl. Phys. A {\bf
553}, 515 (1993).
\bibitem{bhf2}
Kh. S. A. Hassaneen and H. M¡§uther, Phys. Rev. C {\bf 70}, 054308
(2004).
\bibitem{bhf3}
P. Wang, S. -X. Gan, P. Yin, and W. Zuo, Phys. Rev. C {\bf 87},
014328 (2013).
\bibitem{bhf4}
M. Baldo, I. Bombaci, G. Giansiracusa, U. Lombardo, C. Mahaux, and
R. Sartor, Phys. Rev. C {\bf 41}, 1748 (1990); Nucl. Phys. A {\bf
545}, 741 (1992).
\bibitem{EBHF}
W. Zuo, I. Bombaci, and U. Lombardo, Phys. Rev. C {\bf 60}, 024605
(1999).
\bibitem{tbf1}
P. Grang\'{e}, A. Lejeune, M. Martzolff, and J.-F. Mathiot, Phys.
Rev. C {\bf 40}, 1040 (1989).
\bibitem{tbf2}
W. Zuo, A. Lejeune, U. Lombardo, and J.-F. Mathiot, Nucl. Phys. A
{\bf 706}, 418 (2002); Eur. Phys. J. A {\bf 14}, 469 (2002).
\bibitem{k4}
O. Hen, L. B. Weinstein, E. Piasetzky, G. A. Miller, M. M. Sargsian,
and Y. Sagi, Phys. Rev. C {\bf 92}, 045205 (2015).
\bibitem{tan1}
S. Tan, Annals of Physics {\bf 323}, 2952 (2008); {\bf 323}, 2971
(2008); {\bf 323}, 2987 (2008).
\bibitem{tan2}
J. T. Stewart, J. P. Gaebler, T. E. Drake, and D. S. Jin, Phys. Rev.
Lett. {\bf 104}, 235301 (2010).
\bibitem{lin}
H. M\"{u}ther, G. Knehr, and A. Polls, Phys. Rev. C {\bf 52}, 2955
(1995).
\bibitem{clus}
S. Typel, G. Ropke, T. Klahn, D. Blaschke, and H. H. Wolter, Phys.
Rev. C {\bf 81}, 015803 (2010).
\bibitem{ust1}
V. J. Emery, Nucl. Phys. A {\bf 12}, 69 (1959); H. F. Arellano and
J. P. Delaroche, Eur. Phys. J. A {\bf 51}, 7 (2015);
\bibitem{ust2}
W. H. Dickhoff, Phys. Lett. B {\bf 210}, 15 (1988); B. E.
Vonderfecht, C. C. Gearhart, W. H. Dickhoff, A. Polls and A. Ramos,
Phys. Lett. B {\bf 253}, 1 (1991).
\bibitem{yonggc17}
G. C. Yong, Phys. Rev. C {\bf 96}, 044605 (2017).
\bibitem{caibj}
B. J. Cai, B. A. Li, and L. W. Chen, Phys. Rev. C {\bf 94}, 061302
(2016).
\bibitem{fopi}
W. Reisdorf \emph{et al.} (FOPI Collaboration), Nucl. Phys. A {\bf
848}, 366 (2010).
\bibitem{sh1}
X. H. Fan, X. L. Shang, J. M. Dong and W. Zuo, Phys. Rev. C. {\bf
99}, 065804 (2019)
\bibitem{sh2}
X. L. Shang, and W. Zuo, Phys. Rev. C. {\bf 88}, 025806 (2013); X.
L. Shang, P. Wang, P. Yin and W. Zuo, J. Phys. G: Nucl. Part. Phys.
{\bf 42}, 055105 (2015).
\end{thebibliography}
\end{document}